\documentstyle[preprint,eqsecnum,aps,prb]{revtex}

\begin{document}
\title{High magnetic field transport measurement of charge-ordered Pr$_{0.5}$Ca$%
_{0.5}$MnO$_{3}$ strained thin films.}
\author{W. Prellier$\thanks{%
prellier@ismra.fr}^{,1}$, E. Rauwel Buzin$^1$, S. de Brion$^2$, G. Chouteau$%
^2$, Ch. Simon$^1$, B.\ Mercey$^1$ and M.\ Hervieu$^1$}
\address{$^1$Laboratoire CRISMAT, CNRS\ UMR 6508, Bd du Mar\'{e}chal Juin, F-14050
Caen\\
Cedex, France\\
$^{2}$Grenoble High Magnetic Field Laboratory, CNRS and MPI-FKF, BP 166,\\
38042 Grenoble Cedex 9, France.}
\date{\today}
\maketitle

\begin{abstract}
We have investigated the magnetic-field-induced phase transition of
charge-ordered (CO) Pr$_{0.5}$Ca$_{0.5}$MnO$_3$ thin films, deposited onto
(100)-oriented LaAlO$_3$ and (100)-oriented SrTiO$_3$ substrates using the
pulsed laser deposition technique, by measuring the transport properties
with magnetic fields up to 22T. The transition to a metallic state is
observed on both substrates by application of a critical magnetic field ($%
H_C>10T$ at $60K$). The value of the field required to destroy the
charge-ordered insulating state, lower than the bulk compound, depends on
both the substrate and the thickness of the film. The difference of the
critical magnetic field between the films and the bulk material is explained
by the difference of in-plane parameters at low temperature (below the CO\
transition). Finally, these results confirm that the robustness of the CO
state, depends mainly on the stress induced by the difference in the thermal
dilatations between the film and the substrate.
\end{abstract}

\bigskip For the past years, there has been a large focus on the charge
ordering (CO) and orbital ordering (OO) phenomena \cite{Tokura} in
transition metal oxides. Interesting compounds showing this behavior include
manganite oxides, such as $Nd_{0.5}Sr_{0.5}MnO_3$ \cite{Mag-Field}, $%
Pr_{1-x}Ca_xMnO_3$ \cite{Tomioka}, or nickelates\ like $La_{2-x}Sr_xNiO_4$ 
\cite{Nickel}. In colossal magnetoresistive (CMR) manganites \cite{Rav},
charge/orbital ordering corresponds to an ordering of the charges/orbitals
in two different Mn sublattices (i.e. a long-range ordering of $Mn^{3+}$ and 
$Mn^{4+}$ ions). It appears for certain value of $x$ and particular average
cation radius. Under cooling, the polaronic transport in the paramagnetic
state becomes unstable, below a certain temperature ($T_{CO}$) and the
material goes to an insulating, charge-ordered (CO) state. $T_{CO}$
decreases with increasing field \cite{CO}. The feature that the
charge-ordering can be destroy leading to metallic-like state, under the
application of an external perturbation like magnetic field \cite
{Mag-Field,Tomioka}, electric field \cite{Elec-Field}, visible-infrared
light \cite{IR}, electron irradiation \cite{ED} or X-rays \cite{X-rays} has
stimulated extensive work with the aim of examining complex structural and
magnetotransport transitions. Interestingly, this fall of resistivity much
larger than the conventional CMR\ materials\cite{CMR}, has increased their
potential use for technologic applications. However, prior to a routine
utilization, we need to correctly control the thin films growth \cite
{Prellier} and their characterizations.

For these reasons, we have first undertaken studies on $%
Pr_{0.5}Ca_{0.5}MnO_3 $ (PCMO) films deposited on $LaAlO_3$ \cite{AMH} and $%
SrTiO_3$ \cite{WilCO1,WilCO2} substrates in order to understand the effect
of substrate-induced strains. While the thinner films do not exhibit any
temperature induced insulator-metal transition under an applied magnetic
field up to $9T$, for thickness larger than $110nm$\ a $5T$ magnetic field
is sufficient to destroy the CO/OO state \cite{WilCO2}. This indicates that
strains play a crucial role in the stability of the CO/OO state. These
previous studies, limited to $9T$, were not very conclusive on the origin of
this effect since the CO/OO state of the thinner films and of the bulk
material do not collapse under a $9T$ magnetic field. Note that a magnetic
field of $25T$ (at $4K$) is required to destroy this insulating state in
bulk $Pr_{0.5}Ca_{0.5}MnO_3$ \cite{Tokunaga}.

In the present work, we have studied the magnetic-field-induced phase
transition of CO/OO $Pr_{0.5}Ca_{0.5}MnO_3$ films in high magnetic fields up
to $22T$. We have carried out transport measurements over a wide temperature
range for two types of samples where any effect was seen up to $9T$: a $25nm$%
\ film grown on $SrTiO_3$ (STO) and a $250nm$\ film deposited on LaAlO$_3$
(LAO) \cite{AMH,WilCO1}. Based on these results, we have determined the
dependence of the critical magnetic field as a function of temperature and
as a function of the film thickness and compared it to what is observed in
the bulk compound. Finally, we have correlated this mechanism with the
structural properties of the thin films.

Thin films of PCMO were grown in-situ using the pulsed laser deposition
technique on (100)-LaAlO$_3$ (pseudocubic with $a=0.3788nm$) and (100)-SrTiO$%
_3$ (cubic with $a=0.3905nm$) substrates. Detailed optimization of the
growth procedure was completed and described previously \cite{AMH,WilCO1}.
The structural study was carried out by X-Ray diffraction (XRD) using a
Seifert XRD 3000P at room temperature (Cu K$\alpha $, $\lambda =0.15406nm$).
Resistivity ($\rho $) was measured by a four-probe method as a function of
the magnetic field ($H$) up to $22T$, for various temperature ($T$) in the
range $4-300K$. The speed of the field used, in ramping up and down, was $%
25mT/sec$ (we also measured the PCMO film on STO with a speed of $55mT/s$
and $85mT/s$ but no changes were observed). The composition of the films was
checked by energy-dispersive spectroscopy analyses. It is homogenous and
corresponds exactly to the composition of the target (i.e.$Pr_{0.5\pm
0.02}Ca_{0.5\pm 0.02}Mn$) in the limit of the accuracy{\it .}

Fig.1 shows a typical $\theta -2\theta $ scan recorded for a film of $25nm$\
film on STO (Fig.1a) and $250nm$\ film on LAO (Fig.1b). As already reported,
the film is a single phase, [010]-oriented (i.e. with the [010] axis
perpendicular to the substrate plane) on STO and [101]-oriented on LAO (i.e.
with the [101] axis perpendicular to the substrate plane in the space group\ 
$Pnma$ \cite{Zirak} and we have attributed this surprising orientation as a
result of the lattice mismatch between the film and the substrate \cite
{WilCO2}. The out-of-plane parameter, at room temperature,\ is $0.376nm$\
for the $25nm$\ film grown on STO substrate and $0.384nm$\ for the $250nm$\
film deposited on LAO substrate confirming that the PCMO film is under
expansion in the plane of STO and under compression in-the-plane of LAO,
respectively.

The magnetic-field dependence of the resistivity at various temperatures for 
$Pr_{0.5}Ca_{0.5}MnO_3$ film on STO\ substrate is shown on Fig.2. The
resistivity shows a huge decrease on a logarithmic scale, showing the
transition toward the charge disordered, metallic phase.\ We define the
critical field $H_C$ for this phase transition at the inflection point.
There is a strong hysteresis between the field-increasing ($H_C^{+}$) and
field-decreasing ($H_C^{-})$ behavior$\ $as observed in the bulk material
but the values are very different. For instance, the critical field at 120K
is close to $15T$ for the $25nm$\ film on STO, whereas in the bulk compound, 
$H_C$\ is around $22T$ \cite{Tokunaga}.

The in-plane misfit $\sigma =100*(a_F-a_B)/a_B$ (where $a_F$ is the in-plane
parameter of the film and $a_B$ is the lattice parameter of bulk PCMO) is
calculated at room temperature. $a_B$ corresponds to the $d_{101}$ in the
case of STO and to the $d_{010}$ in the case of LAO respectively, due to the
different orientation of the film with respect to the substrate. The
evolution of the critical magnetic field as a function of this misfit, at
room temperature, is presented Fig.3a. At $T=300K$, there is no clear
correlation between these two parameters. In particular, the value of $H_C$
tends, for the highest thicknesses of the films on STO substrates, to a
value which is completely different from that of the bulk material. We have
already discussed in previous papers the possible explanation in terms of
changes of the composition or in the oxygen content: this explanation was
ruled out completely by the fact that the structural parameters relaxes to
the bulk value when the film is removed from the substrate by scratching 
\cite{AMH}. In particular the modulation vector \cite{Modulation}, $q$, of
the CO/OO state which is $0.48$ in the bulk stoichiometric sample is only $%
0.38$ in one of the film on LAO substrates. After scratching (i.e. when the
film is substrate-free), it comes back to the bulk value of $0.48$ \cite{AMH}
indicating that the value of $0.38$ is only a result of the
substrate-induced strains.

The interpretation that we propose is that, when the film remains epitaxial
on the substrate, the CO state cannot fully develop because it is impossible
to accommodate the quite large change in the structural parameters occurring
below $T_{CO}$\ in the bulk \cite{Martin}. For example, in the bulk PCMO,\
the [101] lattice parameter (which should be compared to the in-plane
parameter on STO) is going from $0.382nm$\ to $0.386nm$\ in the CO state.
Since the low temperature ED study has shown that the film remains epitaxial
below the CO transition, the in-plane lattice parameter of the film cannot
reach $0.386nm$\ and remains close to the substrate value ($0.390nm$) for
the thinner film.

\smallskip For the thicker films, at the synthesis temperature, the in-plane
lattice parameters relax smoothly across its thickness ($t$) from the STO
value to a value $0.381nm$\ close to the bulk one. At room temperature, the
values are slightly smaller but remains almost unchanged ($0.390nm$\ and $%
0.381nm$\ for the substrate and the film respectively). For characterizing
the nanostructural state of the films and, more especially, understanding
the way the relaxation is ensured as the thickness increases, an electron
diffraction (ED) and high resolution electron microscopy (HREM) study was
carried out on the thicker film ($t>200nm$). The study was performed with a
TOPCON 002B electron microscope having a $0.18nm$\ point to point resolution
($Voltage=200kV$ and spherical aberration coefficient $C_S=0.4mm$). The ED
patterns confirm that the whole film is [010]-oriented and evidence also the
existence of twinning domains. Such domains, resulting from the orthorhombic
distortion of the perovskite subcell, have been extensively described \cite
{distorsion}. The original character of the present film, by opposition to
the bulk and certain films \cite{Lebedev}) is that only two variants out of
six are observed, namely those with the [100] and [001] directions parallel
the \{110\}$_{STO}$ equivalent directions. These points are illustrated in
Fig.4a (only one quadrant of the ED pattern is given for allowing a
sufficient enlargement) and in the HREM image in Fig.4c. These ED patterns
provide two other important informations. First, the $600$ and $006$
reflections of the film are perfectly superimposed showing that the $a$ and $%
c$ parameters are equal. The through focus HREM series confirmed the
homogeneity of the film structure (in agreement with the simulated images,
calculated with a Mc TEMPAS software). Second, the conditions of reflection (%
$Pnma$ space group) show that the symmetry of the cell remains orthorhombic
despite this particular geometrical relationship imposed by the substrate,
i.e. the tilting mode of the octahedra is similar to that of the bulk
material.

The overall images of the film (Fig.4b) show alternating broad dark and
bright bands perpendicular to the substrate plane, which are characteristic
of strain effects (and associated to the relaxation mode). The evolution of
the $a$, $b$ and $c$ parameters throughout the films, was determined by
measuring directly on the HREM images, taking those of the substrate as
references \cite{TEM}. It shows that, concomitantly and continuously, $a$
(and $c$, since $a=c$) increases whereas $b$ decreases when going away from
the interface. This evolution is made at roughly constant cell volume (in
the limit of accuracy of the measurements). Large areas of the film were
carefully investigated. No dislocation has been detected, whatever the film
zone, ruling out definitely such a structural mechanism for explaining the
relaxation. The detailed examination (reported elsewhere) allows a mechanism
of smooth variation to be proposed. The images indeed show very local
variations of the contrast. They appear as point like defects and waving
atomic rows, associated to ion displacements and local distortions of the
octahedra. This is exemplified in film areas very close to the interface
(circled in Fig. 4c); the effect is clearly visible by viewing at grazing
incidence. These phenomena are observed in the whole film and generate a
tiny mosaicity of the film, responsible of the stripe-like contrast in the
images (Fig.4b).

In conclusion, the electron microscopy study showed that the film exhibits
the same GdFeO$_3$-type structure (orthorhombic-$Pnma$ space group) as the
corresponding bulk material but with different lattice parameters and,
consequently, different inter-atomic distances and inter-bond angles (in
particular Mn-O distances and Mn-O-Mn angles). This effect is moreover
accentuated by the strain effects, which are directly correlated to the
in-plane misfit.

One would thus expect that the effect of the in-plane mismatch is more
crucial below the CO temperature than at room temperature. Thus, we did the
calculations of the mismatch at $120K$, a temperature below $T_{CO}$. For
this, we consider that the in-plane parameters of the films have a tiny
variation under cooling when going from $300K$ to $120K$ since the lattice
parameters of the substrate is almost constant \cite{STO}. The resulting
graph, calculated for STO substrates, is presented in Fig.3b. When the
misfit is equal to zero (corresponding to the bulk value), the critical
field is around $20T$ \cite{Tokunaga}. It appears to be a maximum since a
value of $-0.5$ (corresponding to a $110nm$\ film) leads to a $H_C$ of $5T$\
whereas a value of $+0.5$ (corresponding to a $25nm$\ film) gives a $H_C$\
of $17T$. We also add in this graph the datas of LAO\ substrate calculated
in the same way. However, it seems difficult to compare the results on LAO
with STO\ since the films behave differently (nature of the strains,
orientation of the film with respect to the substrate).

A $250nm$\ thick film of PCMO grown on LAO was also investigated. Fig.5
shows the magnetic-field dependence of the resistivity at various
temperatures. At $120K$, $H_C$\ is close to $10T$ which is also much lower
than the bulk value. In order to compare these datas with films grown on
STO, we are currently undertaken studies of films on LAO with various
thickness.

One remark should also be made. $H_{C}$ is always smaller in the case of the
thin film. This means that the nature of the strains (i.e. compression or
expansion) always drives $H_{C}$ in the same direction due to the fact that
the CO/OO\ state is less stable in a thin film. In others words, the CO/OO\
state is less established and it is easier to collapse it. This can be
explained regarding the orientation of the film. On STO, the [010]-axis
(corresponding to the out-of-plane direction) is compressed whereas the
[101]-axis (corresponding to the in-plane direction) is expanded. On LAO,
the [101]-axis is also compressed but this effect occurs in-the-plane of the
substrate. Thus, the expansion of the [010]-axis destabilizes the
orbital-ordering phase as seen in bulk $Nd_{1-x}Sr_{x}MnO_{3}$ where a
compression of the [010]-axis assists the cooperative Jahn-Teller distortion
and stabilizes the orbital-ordering state \cite{Arima}. The difference of
the $H_{C}$ values between LAO and STO\ is due to the different mismatches
between both substrates but the effect is similar.

In conclusion, the main parameter to control the CO/OO state is the low
temperature in-plane parameter. When it reaches a value far from the bulk
value, the CO/OO state cannot fully develop. This can be seen on the
modulation vector parameter as it was previously published, but also, as it
is found here, on the stability energy of the phase which can be directly
calculated from the critical field (by multiplication by the moment of the
ions). On STO substrate, the compression of the [010]-axis is measured. On
LAO substrate, the same effect is observed along the [101]-axis.

We acknowledge Prof. B. Raveau, Dr. A. Maignan and Dr. A. Ambrosini for
fruitful discussions and carefull reading of the article.

Figure Captions

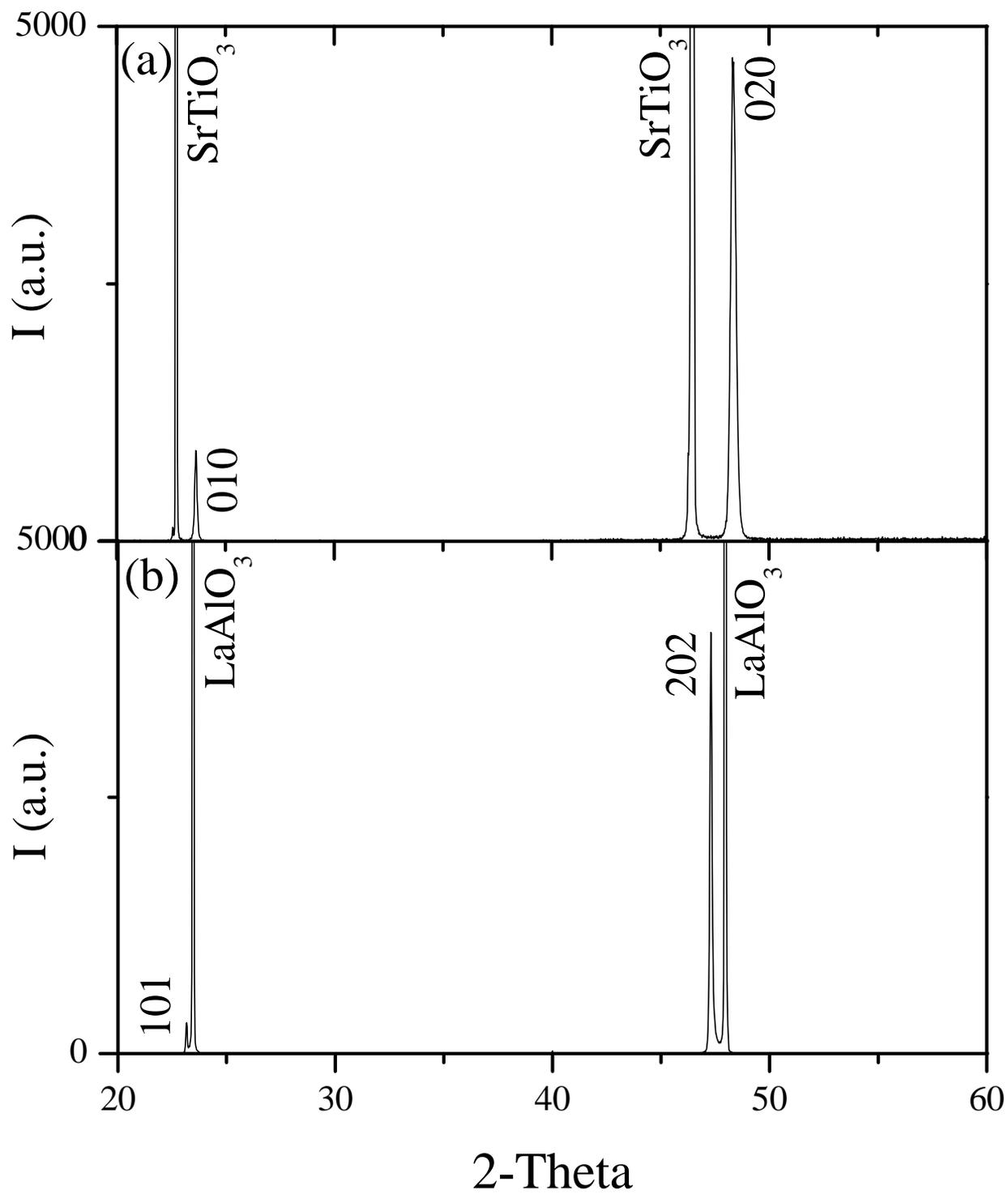
Fig.1 : Room temperature $\theta -2\theta $ XRD pattern of a PCMO thin\ film
(a): $25nm$\ on SrTiO$_3$, (b): $250nm$\ on LaAlO$_3$.

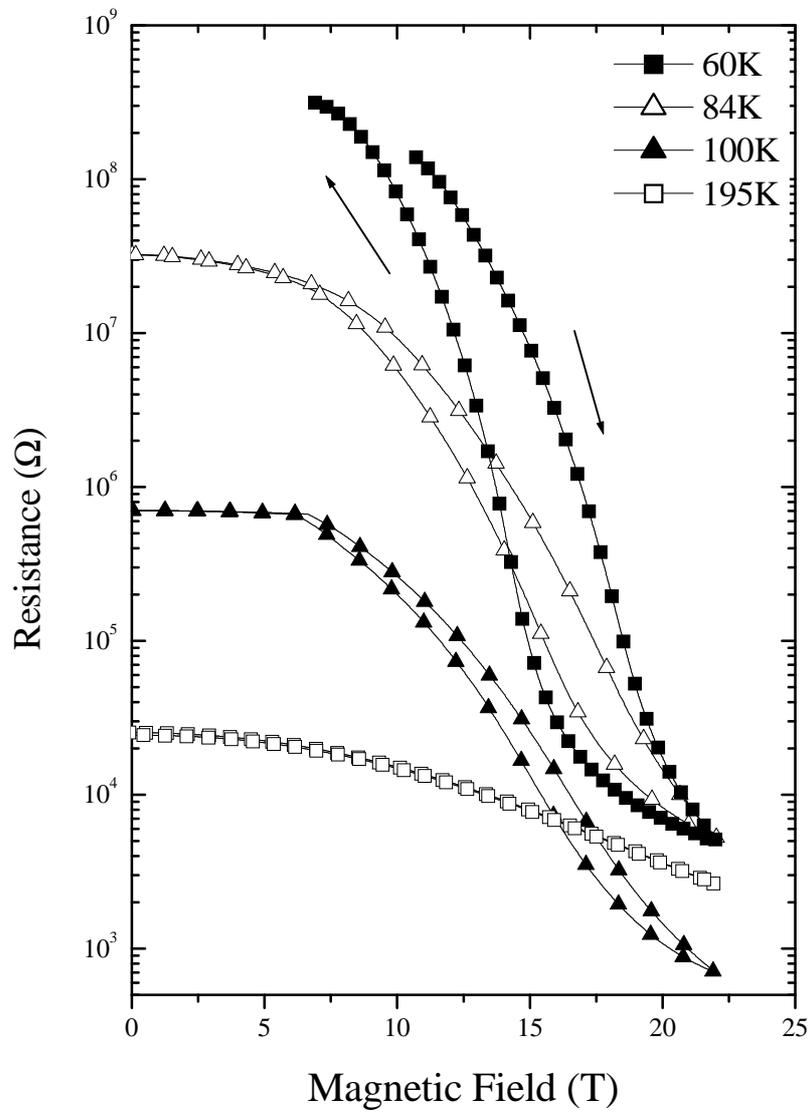
Fig.2: $\rho (H)$ at different temperatures for a $25nm$\ PCMO thin films
grown on SrTiO$_3$. \ Runs in field increasing and decreasing are indicated
by arrows.

Fig.3: Evolution of the critical magnetic field taken at $120K$, as a
function of\ the in-plane misfit for different films (squares: STO,
triangles: LAO) calculated (a): at room temperature (full symbols) and (b):
below $T_{CO}$ (empty symbols). Note the shift of the line at zero
(corresponding to the bulk material) between $T=300K$ and $T<T_{CO}$. The
dashed and dots lines are only a guide for the eyes (see text for details).

Fig. 4a: ED of a cross-section for a PCMO/STO showing the [010]-axis
perpendicular to the substrate plane. The spots allow an orthorhombic
symmetry of the film, as seen in the bulk, but since the $600$ and $006$
reflections of the film are perfectly superimposed; this results in $a=c$.
Subscripts F\ and S correspond to the substrate and the film respectively.

Fig. 4b: Overall cross-section image (bright field) of a PCMO film on STO
showing the contrast typical of strain effects.

Fig. 4c: Cross-section (bright field) HREM image taken close to the
interface film/substrate. No change of the lattice parameters are visible at
the interface (marked by white arrows). The white circles (one is
exemplified on the top of the image) show a local variation of the contrast.
This variations indicate local distortions of the cell and waving atomic
rows resulting of the smooth relaxation of the films due to the strains.
Subscripts F\ and S correspond to the substrate and the film respectively.

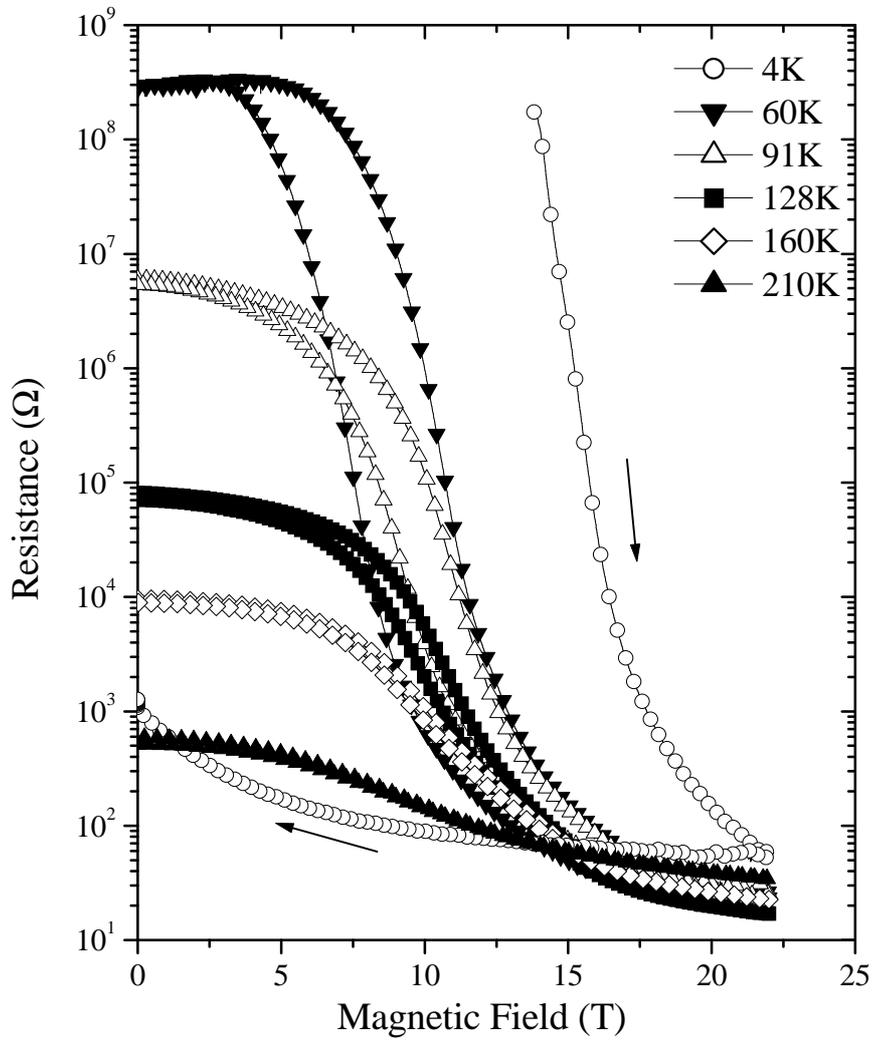
Fig.5: $\rho (H)$ at different temperatures for $250nm$\ PCMO thin films
grown on LaAlO$_3$.\ Runs in field increasing and decreasing are indicated
by arrows.


\begin{references}
\bibitem{Modulation}  The charge-orbital modulation vector $q$ exhibits a
value of $q=1/2-\epsilon $ where \ the incommensurability ($\epsilon $) is
zero for a perfect 1:1 charge-ordering.
\bibitem{TEM}  The STO substrate was used as a reference by measuring the
distance between two dots, which corresponds a known unit cell ($0.3905nm$),
away from the interface to avoid any local distortion . The lattice
parameters of the film were measured in the same way, taking the distance
between two dots which corresponding to a known integral number of unit
cells.
\end{references}
\end{document}